# Generalized Discriminant Analysis algorithm for feature reduction in Cyber Attack Detection System

Shailendra Singh
Department of Information Technology
Rajiv Gandhi Technological University
Bhopal, India

Sanjay Silakari
Department of Computer Science and Engineering
Rajiv Gandhi Technological University
Bhopal, India

*Abstract*—This Generalized Discriminant Analysis (GDA) has provided an extremely powerful approach to extracting non-linear features. The network traffic data provided for the design of intrusion detection system always are large with ineffective information, thus we need to remove the worthless information from the original high dimensional database. To improve the generalization ability, we usually generate a small set of features from the original input variables by feature extraction. The conventional Linear Discriminant Analysis (LDA) feature reduction technique has its limitations. It is not suitable for non-linear dataset. Thus we propose an efficient algorithm based on the Generalized Discriminant Analysis (GDA) feature reduction technique which is novel approach used in the area of cyber attack detection. This not only reduces the number of the input features but also increases the classification accuracy and reduces the training and testing time of the classifiers by selecting most discriminating features. We use Artificial Neural Network (ANN) and C4.5 classifiers to compare the performance of the proposed technique. The result indicates the superiority of algorithm.

## I. INTRODUCTION

Information assurance is an issue of serious global concern. The internet has brought about great benefits of the modern society. According to the statistics of American Computer Emergency Response Team /Coordination Center (CERT) [1], network cases annually showed index growth in recent years and according to the report of information security [2] internet attacks have became new weapon of world war. Further the report said that Chinese Military Hacker had drew up plan, with the view of attacking American Aircraft Carrier Battle Group to making in it weak fighting capacity thorough internet. Such information reveals that there is an urgent need to effectively identify and hold up internet attacks. It is not an exaggerated statement that an intrusion detection system is must for modern computer systems. Anomaly detection and misuse detection [3] are two general approaches to computer intrusion detection system. Unlike misuse detection, which generates an alarm when a known attack signature is matched, anomaly detection identifies activities that deviate from the normal behavior of the monitored system and thus has the potential to detect novel attacks [4]. Currently there are three basic approaches [5] for cyber attack detection. The data we use here originated from MIT's Lincoln Lab. It was developed for KDD (Knowledge Discovery and Data mining) competition by DARPA and is considered a standard benchmark for intrusion detection evaluation program [6]. Empirical studies indicate that feature reduction technique is capable of reducing the size of dataset. The time and space complexities of most classifiers used are exponential function of their input vector size [7]. Moreover, the demand for the number of samples for the training the classifier grows exponentially with the dimension of the feature space. This limitation is called the 'curse of dimensionality.'

The feature space having reduced features that truly contributes to classification that cuts pre-processing costs and minimizes the effects of the 'peaking phenomenon' in classification [8]. Thereby improving the over all performance of classifier based intrusion detection systems. The most famous technique for dimensionality reduction is Linear Discriminant Analysis [9] [10]. This technique searches for directions in the data that have largest variance and subsequently project the data into it. By this we obtain a lower dimensional representation of the data that removes some of the "noisy" directions. But this suffers from many difficult issues with how many directions one needs to choose. It fails to compute principal component in high dimensional feature spaces, which are related to input space by some nonlinear map.

In this paper we present Generalized Discriminant Analysis (GDA) [11] technique to overcome the limitations of LDA technique. This is unique approach to reduced size of attack data The Each network connection is transformed into an input data vector. GDA is employed to reduce the high dimensional data vectors and identification is handled in a low dimensional space with high efficiency and low use of system resources. The normal behavior is profiled based on normal data for anomaly detection and the behavior of each type of attack are built based on attack data for intrusion identification. Each reduced feature dataset is applied to the Artificial Neural

Network (ANN) and C4.5 decision tree classifiers and their performance are compared.

## II. THE DATA SET

In the 1998 DARPA intrusion detection evaluation program [6], an environment was setup to acquire raw TCP/IP dump data for a network by simulating a typical U.S. Air Force LAN. The LAN was operated like a true environment, but being blasted with multiple attacks. For each TCP/IP connection, 41 various quantitative (continuous data type) and qualitative (discrete data type) features were extracted among the 41 features, 34 features are numeric and 7 features are symbolic. The data contains 24 attack types that could be classified into four main categories:

- DOS: Denial of Service attack.
- R2L: Remote to Local (User) attack.
- U2R: User to Root attack.
- Probing: Surveillance and other probing.

### A. Denial of service Attack (DOS)

Denial of service (DOS) is class of attack where an attacker makes a computing or memory resource too busy or too full to handle legitimate requests, thus denying legitimate user access to a machine.

### B. Remote to Local (User) Attacks

A remote to local (R2L) attack is a class of attacks where an attacker sends packets to a machine over network, then exploits the machine's vulnerability to illegally gain local access to a machine.

### C. User to Root Attacks

User to root (U2R) attacks is a class of attacks where an attacker starts with access to a normal user account on the system and is able to exploit vulnerability to gain root access to the system.

### D. Probing

Probing is class of attacks where an attacker scans a network to gather information or find known vulnerabilities. An attacker with map of machine and services that are available on a network can use the information to notice for exploit.

## III. TECHNIQUES FOR FEATURE EXTRACTION

Feature extraction applies a mapping of the multidimensional space into a space of lower dimensions. Feature extraction [12] includes feature construction, space dimensionality reduction, sparse representations, and feature selection. All these techniques are commonly used as pre processing to machine learning and statistics tasks of prediction, including pattern recognition and regression. Although such problems have been tackled by researchers for many years, there has been recently a renewed interest in feature extraction. A number of new applications with very large input spaces critically need space dimensionality reduction for efficiency of the classifiers.

In this section we discuss two techniques LDA and proposed GDA for reducing dimensionality of KDDCup99 intrusion detection dataset. Each feature vectors is labeled as an attack or normal. The distance between a vector and its reconstruction onto those reduced subspaces representing different types of attacks and normal activities is used for identification.

### A. Linear Discriminant Analysis (LDA)

Linear Discriminant Analysis [9][10][13] is a class specific method in the sense that it represents data to make if useful for classification. Finds the optimal transformation matrix as to preserve most of the information that can be used to discriminate between the different classes. Therefore the analysis requires the data to have appropriate class labels. In order to mathematically formulate the optimization

Let $X = \{x_1, x_2, ......x_M\}$ be the dataset given N-dimensional vectors of KDDCup99 dataset. Each data point belongs to one of C object classes $\{X_1, X_2, ......X_C\}$. The between class scatter matrix and the within-class scatter matrix are defined as

$$B = \sum_{c=1}^{C} M_c (m_c - m)(m_c - m)^T \qquad (1)$$

$$W = \sum_{c=1}^{C} \sum_{x \in X_{cc}} (x_c - m_c)(x - m_c)^T \qquad (2)$$

Where $m_c$ denotes the class mean and m is the global mean of the entire sample. The number of vectors in class $x_c$ is denoted by $M_c$. LDA finds matrix, U, maximizing the ratio of determinant of the between-class scatter matrix to the determinant of the within-class scatter matrix a

$$U_{opt} = \arg\max U \frac{|(U^T BU)|}{|(U^T WU)|} = [u_1, u_2 ..... u_N]. \qquad (3)$$

The solution $\{u_i | i = 1,2,3......N\}$ is a set of generalized eigenvectors of B and W, i.e. $Bu_i = \lambda_i W u_i$.

With these definitions, we can easily formulate the optimization criterion. Namely the numerator represents the covariance of the pooled training data in the transformed feature space. The denominator represents the average covariance within each class in the transformed feature space. Hence, the criterion really tries to maximize the 'distance' between classes, while minimizing the 'size' of each of the classes at the same time. This is exactly what we want to achieve because this criterion guarantees that we preserve most of the discriminant information in the transformed feature space. It turns out that the optimum matrix according to the above formula can be found in a fairly easy way. LDA is applied to the KDDCUP99 data and the features selected are given below

TABLE I. FEATURES SLECTED BY LDA TECHNIQUE

| S. No | Feature | Type |
|---|---|---|
| 1 | duration | Continuous |
| 2 | protocol_type | Discrete |
| 3 | service | Discrete |
| 4 | src_bytes | Continuous |
| 5 | land | Discrete |
| 6 | wrong_fragment | Continuous |
| 7 | num_failed_logins | Continuous |
| 8 | logged_in | Discrete |
| 9 | root_shell | Continuous |
| 10 | num_file_creation | Continuous |
| 11 | is_guest_login | Discrete |
| 12 | count | Continuous |
| 13 | srv_count | Continuous |
| 14 | serror_rate | Continuous |
| 15 | srv_serror_rate | Continuous |
| 16 | diff_srv_rate | Continuous |
| 17 | dst_host_count | Continuous |

TABLE II. CONFUSION MATRIX FOR ANN CLASSIFIER BY LDA TECHNIQUE.

| Predicted / Actual | Normal | Probe | DOS | R2L | U2R | %Correct |
|---|---|---|---|---|---|---|
| Normal | 58748 | 773 | 1070 | 1 | 1 | 96.95 |
| Probe | 104 | 40002 | 59 | 1 | 0 | 96.06 |
| DOS | 4211 | 2805 | 222833 | 1 | 3 | 96.94 |
| R2L | 13359 | 1550 | 474 | 180 | 1 | 10.4 |
| U2R | 57 | 127 | 4 | 0 | 40 | 17.54 |
| %Correct | 76.81 | 43.23 | 99.28 | 99.83 | 88.88 | |

TABLE III. CONFUSION MATRIX FOR C4.5 CLASSIFIER BY LDA TECHNIQUE.

| Predicted / Actual | Normal | Probe | DOS | R2L | U2R | %Correct |
|---|---|---|---|---|---|---|
| Normal | 59969 | 423 | 190 | 5 | 6 | 98.17 |
| Probe | 194 | 3881 | 90 | 1 | 0 | 93.15 |
| DOS | 17927 | 8969 | 202942 | 10 | 5 | 88.29 |
| R2L | 13813 | 614 | 6 | 1726 | 30 | 22.3 |
| U2R | 149 | 20 | 2 | 6 | 51 | 10.66 |
| %Correct | 65.14 | 27.90 | 99.85 | 98.74 | 52.43 | |

*B. Generalized Discriminant Analysis (GDA)*

The Generalized Discriminant Analysis is used for multi-class classification problems. Due to the large variations in the attack patterns of various attack classes, there is usually a considerable overlap between some of these classes in the feature space. In this situation, a feature transformation mechanism that can minimize the between-class scatter is used.

The Generalized Discriminant Analysis GDA [11][14] is a method designed for nonlinear classification based on a kernel function $\phi$ which transform the original space X to a new high-dimensional feature space $Z: \phi: X \rightarrow Z$. The within-class scatter and between-class scatter matrix of the nonlinearly mapped data is

$$B^\phi = \sum_{c=1}^{C} M_c m_c^\phi (m_c^\phi)^T \quad (4)$$

$$W^\phi = \sum_{c=1}^{C} \sum_{x \in X_c} \phi(x)\phi(x)^T \quad (5)$$

Where $m_c^\phi$ is the mean of class $x_c$ in Z and $M_c$ is the number of samples belonging to $x_c$. The aim of the GDA is to find such projection matrix $U^\phi$ that maximizes the ratio

$$U_{opt}^\phi = \arg\max \frac{|(U^\phi)^T B^\phi U^\phi|}{|(U^\phi)^T W^\phi U^\phi|} = [u_1^\phi,....,u_N^\phi] \quad (6)$$

The vectors, $u^\phi$, can be found as the solution of the generalized eigenvalue problem i.e. $B^\phi u_i^\phi = \lambda_i W^\phi u_i^\phi$. The training vectors are supposed to be centered (zero mean, unit variance) in the feature space Z. from the theory of reproducing kernels any solution $u^\phi \in Z$ must lie in the span of all training samples in Z, i.e.

$$u^\phi = \sum_{c=1}^{C} \sum_{i=1}^{M_c} \alpha_{ci} \phi(x_{ci}) \quad (7)$$

Where $\alpha_{ci}$ are some real weights and $x_{ci}$ is the $i$th sample of the class c. The solution is obtained by solving

$$\lambda = \frac{\alpha^T KDK \, \alpha}{\alpha^T KK \, \alpha} \quad (8)$$

Where $\alpha=(\alpha_c)$ c=1...C is a vector of weights with $\alpha=(\alpha_{ci}), i=1...M_c$. The kernel matrix $K(M \times M)$ is composed of the dot products of nonlinearly mapped data, i.e.

$$K = (K_{kl})_{k=1...C, l=1....C} \quad (9)$$

Where $K_{kl}=(k(x_{ki}x_{lj}))_{i=1...M_k, j=1....M_l}$ The matrix $D(M \times M)$ is a block diagonal matrix such that

$$D = (D_c)_{c=1...C} \quad (10)$$

Where the $c^{th}$ on the diagonal has all elements equal to $1/M_c$. Solving the eigenvalue problem yields the coefficient vector $\alpha$ that define the projection vectors $u^\phi \in Z$. A projection of a testing vector $x_{test}$ is computed as

$$(u)^T \phi(x_{test}) = \sum_{c=1}^{C} \sum_{i=1}^{M_c} \alpha_{ci} k(x_{ci}, x_{test}) \quad (11)$$

The procedure of the proposed algorithm could be summarized as follows:

- Compute the matrices K and D by solving the equation(9) and(10),
- Decompose K using eigenvectors decomposition,
- Compute eigenvectors $\alpha$ and eigenvalues of the equation(6),
- Compute eigenvectors $u^\phi$ using $\alpha_{ci}$ from equation (7) and normalize them,
- Compute projections of test points onto the eigenvectors $u^\phi$ from equation (11).

The input training data is mapped by a kernel function to a high dimensional feature space, where different classes is supposed to be linearly separable. The Linear Discriminant Analysis (LDA) [15] scheme is then applied to the mapped data, where it searches for those vectors that best discriminate among the classes rather than those vectors that best describe the data [16]. Furthermore, gives a number of independent features which describe the data, LDA creates a linear combination of the features that yields the largest mean differences to the desired classes [17] The number of original 41 features is reduced to 12 features by GDA as shown in the Table IV.

TABLE IV. FEATURES SELECTED BY GENERALIZED DISCRIMINANT ANALYSIS

| S.No | Feature | Type |
|---|---|---|
| 1 | Service | Discrete |
| 2 | src_bytes | Continuous |
| 3 | dst_bytes | Continuous |
| 4 | logged_in | Discrete |
| 5 | Count | Continuous |
| 6 | srv_count | Continuous |
| 7 | serror_rate | Continuous |
| 8 | rv_rerror_rate | Continuous |
| 9 | srv_diff_host_rate | Continuous |
| 10 | dst_host_count | Continuous |
| 11 | dst_host_srv_count | Continuous |
| 12 | dst_host_diff_srv_rate | Continuous |

TABLE V. CONFUSION MATRIX FOR ANN CLASSIFIER BY GDA TECHNIQUE.

| Predicted Actual | Normal | Probe | DOS | R2L | U2R | %Correct |
|---|---|---|---|---|---|---|
| Normal | 59975 | 430 | 192 | 5 | 6 | 98.95 |
| Probe | 100 | 4010 | 55 | 0 | 1 | 96.25 |
| DOS | 2585 | 552 | 226710 | 4 | 2 | 98.63 |
| R2L | 11562 | 3027 | 8 | 1956 | 1 | 12.08 |
| U2R | 99 | 67 | 8 | 1 | 55 | 24.12 |
| %Correct | 69.83 | 25.1 | 99.7 | 99.6 | 76.3 | |

TABLE VI. CONFUSION MATRIX FOR C4.5 CLASSIFIER BY GDA TECHNIQUE..

| Predicted Actual | Normal | Probe | DOS | R2L | U2R | %Correct |
|---|---|---|---|---|---|---|
| Normal | 60400 | 151 | 38 | 1 | 3 | 99.68 |
| Probe | 10 | 4150 | 4 | 1 | 1 | 99.61 |
| DOS | 3058 | 160 | 227339 | 2 | 3 | 98.60 |
| R2L | 3468 | 984 | 1010 | 10726 | 1 | 66.25 |
| U2R | 46 | 47 | 4 | 1 | 130 | 57.01 |
| %Correct | 90.17 | 75.56 | 99.53 | 99.95 | 94.2 | |

TABLE VII. SUMMARY OF DATASET OBTAINED AFTER FEATURE EXTRACTION

| Dataset Name | Features | Method |
|---|---|---|
| ORIGDATA | 41 | None |
| LDADATA | 17 | LDA |
| GDADATA | 12 | GDA |

The resulting confusion matrices of ANN and C4.5 classifiers are obtained as shown in the Table V and VI respectively. We obtain two reduced datasets by LDA and GDA techniques in addition to the original dataset as shown in Table VII.

IV. EXPERIMENTAL RESULT

We will conduct two experiments one with Artificial Neural Network (ANN) [18] and another with C4.5 [19] for training and testing. There are approximately 4,94,020 kinds of data in training dataset and 3,11,029 kinds of data in test dataset of five classes (Normal, DOS,R2L,U2R and Probe). We choose 97277, 391458, 1126, 52 an d 4107 samples for Normal, DOS, R2L, U2R and Prob respectively to train the PCA and proposed GDA and then used test data 60593, 229853, 16189, 228, and 4166 for Normal, DOS, R2L, U2R and Prob respectively to compare the training and testing time and recognition rate. Each sample vector is of dimensionality 41. We use Gaussian kernel $k(x, y) = \exp(-\|x - y\|^2 / 0.1)$ to calculate the kernel matrix. All these experiments are run on the platform of Windows XP with 2.0GHz CPU and 1GB RAM by Weka3.5.8 software to implement the proposed technique.

*A. Artificial Neural Network (ANN)*

We use Artificial Neural Network (ANN) for classification of cyber attacks. In this we use multi-layer feed forward neural network. Since a (multi-layer feed forward) ANN is capable of making multi-class classification. A single ANN is employed to perform the cyber attack detection, using same training and testing sets as those for C4.5. ANN takes long time to train or fail to converge at all when the number of patterns gets large.

*B. C4.5 classifier*

Algorithms for constructing decision trees are among the most well known and widely used of all machine learning methods. Among decision tree algorithms, J. Ross Quinlan's ID3 and its successor, C4.5, are probably the most popular in the machine learning community. These algorithms and variations on them have been the subject of numerous research papers since Quinlan introduced ID3. Classification tree is a prediction mode in machine learning and it is also called Decision tree. It is tree pattern graph similar to flow chart structure; any internal nodes of leaves represent distributed situation of various types. There are two methods for tree construction; top-down tree construction and bottom-up pruning, C4.5 used top-down tree construction.

The detection and identification of attack and non-attack behaviors can be generalized as follows:

*True Positive* (TP): the amount of attack detected when it is actually attack.

*True Negative* (TN): the amount of normal detected when it is actually normal.

*False Positive* (FP): the amount of attack detected when it is actually normal (False alarm).

*False Negative* (FN): the amount of normal detected when it is actually attack.

Confusion matrix contains information actual and predicted classifications done by a classifier. In the performance of such a system is commonly evaluated using the data in a matrix. Table VIII shows the confusion matrix.

TABLE VIII. CONFUSION MATRIX

| Predicted Actual | Normal | Attack |
|---|---|---|
| Normal | True Negative (TN) | False Pasitive (FP) |
| Attack | False Negative (FN) | True Positive (TP) |

In the confusion matrix above, rows correspond to predicted categories, while columns correspond to actual categories.

**Comparison of detection rate:** Detection Rate (DR) is given by.

$$DR = \frac{TP}{TP + FN} \times 100 \%$$

**Comparison of false alarm rate:** False Alarm Rate (FAR) refers to the proportion that normal data is falsely detected as attack behavior.

$$FAR = \frac{FP}{FP + TN} \times 100 \%$$

The reported results in term of detection rate, false alarm rate, training time and testing time of ANN and C4.5 decision tree classifiers are summarized in Tables IX, X.

TABLE IX. DETECTION RATE, FALSE ALARM RATE, TRAINING TIME AND TESTING TIME OF ANN AND C4.5 CLASSIFIER WITH LDA TECHNIQUE

|  | ANN | | | | C4.5 | | | |
|---|---|---|---|---|---|---|---|---|
|  | DR | FAR | TR. | TE. | DR | FAR | TR. | TE. |
| Normal | 96.95 | 23.19 | 44s | 31s | 98.17 | 34.86 | 41s | 30s |
| Probe | 96.15 | 56.77 | 16s | 15s | 93.15 | 72.1 | 16s | 16s |
| DOS | 96.94 | 0.72 | 55s | 27s | 88.29 | 0.15 | 51s | 27s |
| R2L | 10.4 | 0.17 | 17s | 15s | 10.66 | 1.26 | 15s | 12s |
| U2R | 17.54 | 11.12 | 10s | 10s | 22.3 | 47.57 | 10s | 9s |

*DR-detection rate, FAR-false alarm rate, TR- training, TE-testing time*

TABLE X. DETECTION RATE, FALSE ALARM RATE, TRAINING TIME AND TESTING TIME OF ANN AND C4.5 CLASSIFIER WITH GDA.TECHNIQUE

|  | ANN | | | | C4.5 | | | |
|---|---|---|---|---|---|---|---|---|
|  | DR | FAR | TR. | TE. | DR | FAR | TR. | TE. |
| Normal | 98.95 | 30.17 | 39s | 25s | 99.68 | 9.83 | 32s | 23s |
| Probe | 96.25 | 74.9 | 15s | 13s | 99.61 | 24.44 | 13s | 11s |
| DOS | 98.63 | 0.3 | 49s | 24s | 98.60 | 0.47 | 45s | 22s |
| R2L | 12.08 | 0.4 | 14s | 11s | 66.25 | 0.05 | 12s | 9s |
| U2R | 24.12 | 23.7 | 10s | 8s | 57.01 | 5.8 | 7s | 6s |

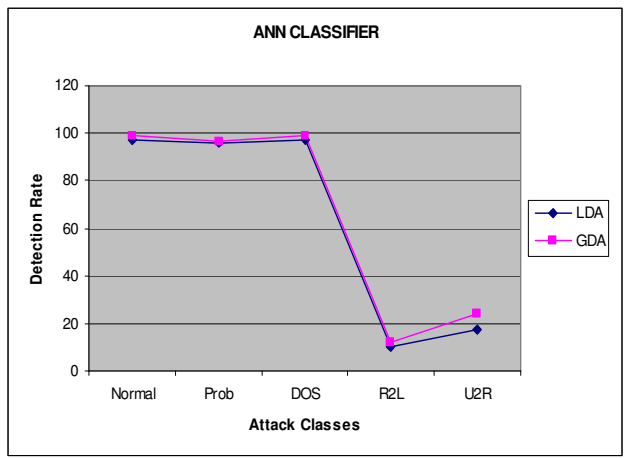

Figure 1. Comparision of detection rate of LDA and GDA for ANN

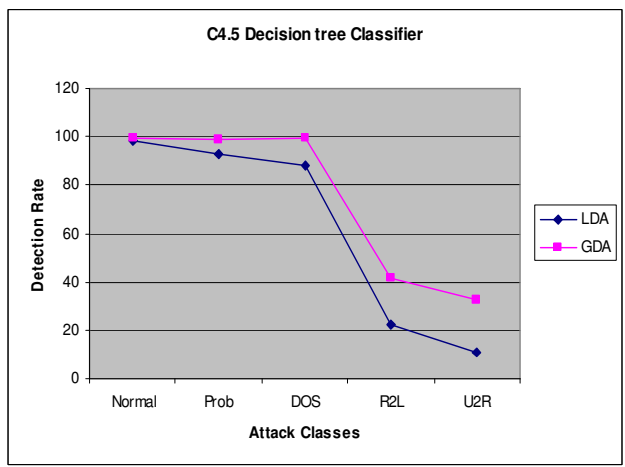

Figure 2. Comparision of detection rate of LDA and GDA for C4.5

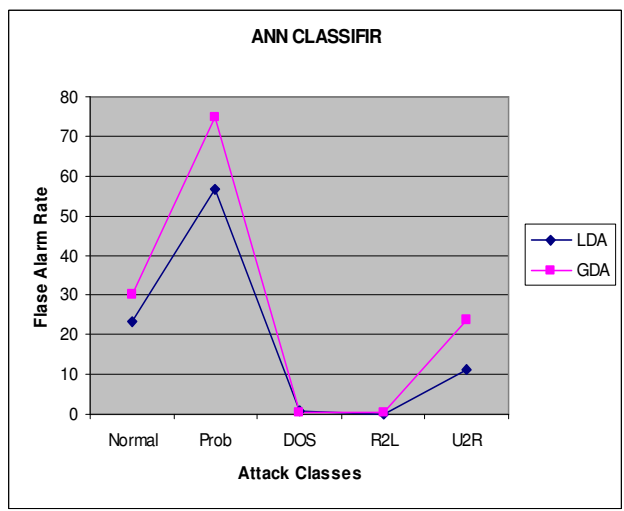

Figure 3. Comparision of false alarm rate of LDA and GDA for ANN

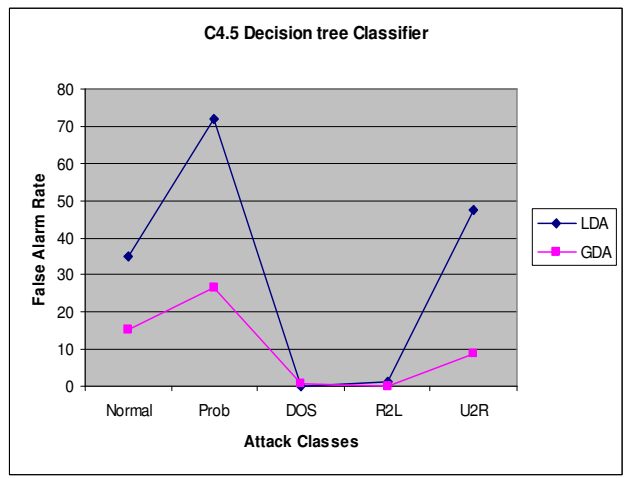

Figure 4. Comparision of false alarm of LDA and GDA for C4.5

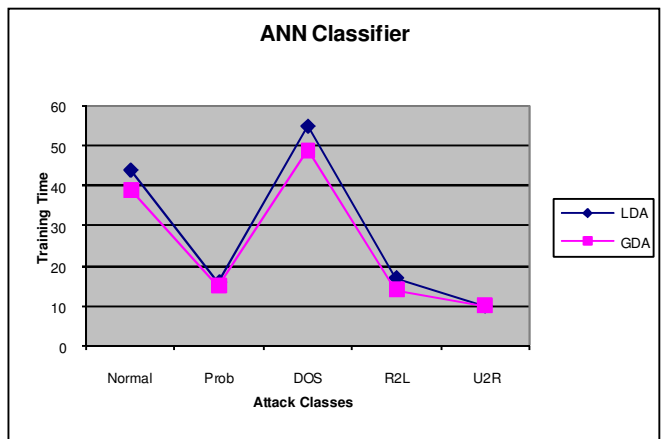

Figure 5. Comparision of training time of LDA and GDA for ANN

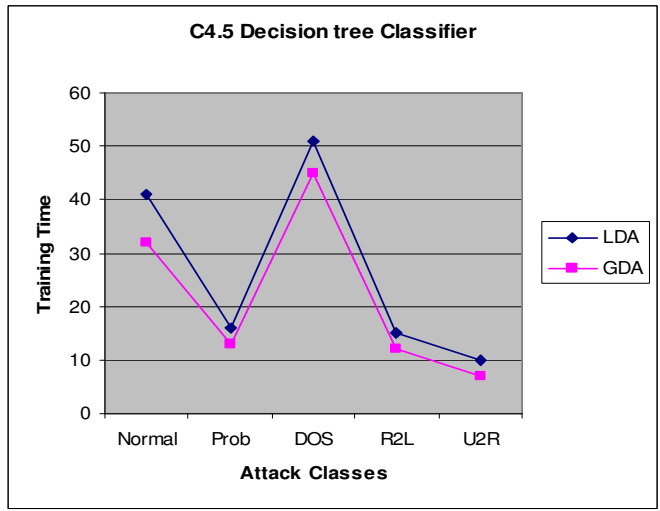

Figure 6. Comparision training time of LDA and GDA for C4.5

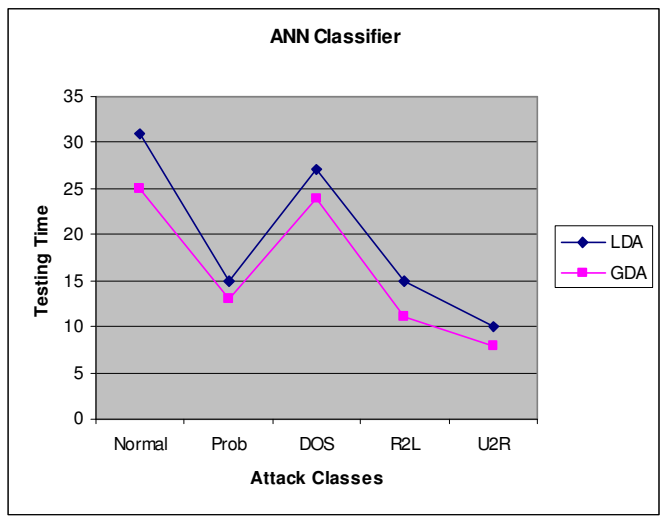

Figure 7. Comparision of testing time of LDA and GDA for ANN

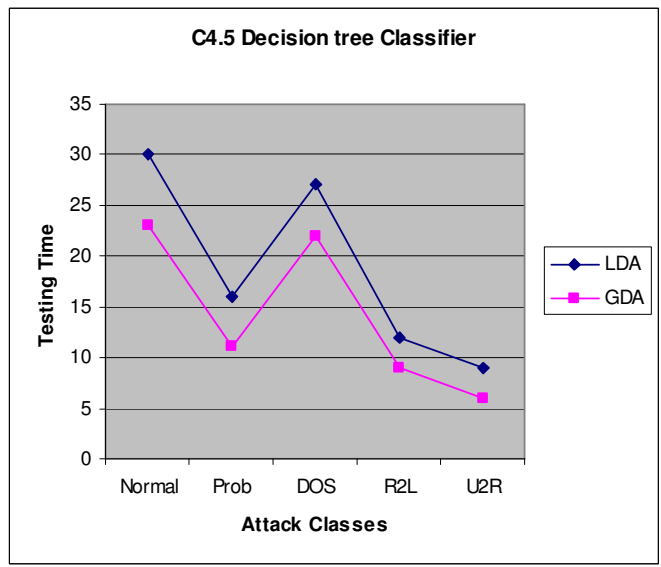

Figure 8. Comparision testing time of LDA and GDA for C4.5

## V. CONCLUSION

As we seen from the result the Generalized Discriminant Analysis algorithm is better than the Liner Discriminant Analysis for the case of large scale dataset where the number of training samples is large. GDA gives better detection rate, less false positives, reduced training and reduced testing times than LDA for the both classifiers. Moreover, when we compared two classifiers, the C4.5 classifier shows better performance for all the classes (Normal, DOS, R2L, U2R, Prob,) and comparables training and testing times as shown in Table IX and X.

Dataset KDDCup99 applied in the research paper is popularly used in current cyber attack detection system; however, it is data of 1999 and network technology and attack methods changes greatly, it can not reflect real network situation prevailing nowadays. Therefore, if newer information is available and tested and compared, they can more accurately reflect current network situation.

We propose ensemble approach for cyber attack detection system in which Generalized Discriminant Analysis (GDA) is used as feature reduction technique and C4.5 as classifier for future research.

REFERENCES

[1] American Computer Emergency Response Team /Coordination Center (CERT),http://www.cert.org/.
[2] Information Security Report, http://www.isecu-tech.com.tw/.
[3] Bace, R.G.: Intrusion Detection. Macmillan Technical Publishing. 2000.
[4] H. Debar etal. "Towards a taxonomy of intrusion detection systems" Computer Network,pp.805-822, April1999.
[5] Shailendra Singh, Sanjay Silakari, " A survey of Cyber Attack Detection Systems" International Journal of Computer Science and Network Security(IJCSNS) Vol.9 No.5.pp.1-10, May,2009.
[6] KDDCup99dataset,August2003 http://kdd.ics.uci.edu/databases/kddcup99/kddcup99.html.
[7] R.O.Duda, P.E.Hart, and D.G.Stork, Pattern Classification, vol. 1. New York: Wiley, 2002.
[8] A.K.Jain, R.P.W.Duin, and J.Mao, "Statistical Pattern Recognition: A Survey," IEEE Transactions on Pattern Analysis and Mission Intelligence, vol. 22, pp.4-37, January 2000. Computer Science
[9] Jing Gao et al."A Novel Framework for Incorporating Labeled Examples into Anomaly Detection", Proceedings of the Siam Conference on Data Mining 2006.
[10] W. Zhao, R. Chellappa, and N. Nandhakumar, "Empirical Performance Analysis of Linear Discriminant Classifiers," Proc.Computer Vision and Pattern Recognition, pp. 164-169, June 1998.
[11] G.Baudt and F. Anouar "Generalized Discriminant Aanlyis Using a Kernal Approach" Neural Computation, 2000
[12] Gopi K. Kuchimanchi,Vir V. Phoha, Kiran S.Balagani, Shekhar R. Gaddam, Dimension Reduction Using Feature Extraction Methods for Real-time Misuse Detection Systems, Proceedings of the IEEE on Information, 2004
[13] Kemal Polat,et.al.. A cascade learning system for classification of diabetes disease: Generalized Discriminant Aanalysis and Least Square Support Vector Machine. Expert Systems with Applications 34 pp-482-487. 2008.
[14] K. Fukunaga. Introduction to Statistical Pattern Classification. Academic Press, San Diego,California, USA, 1990
[15] Kim HC et al. Face recognition using LDA mixture model. In: Proceedings int conf. on pattern recognition, 2002.
[16] Martinez AM, Kak AC. PCA versus LDA. IEEE Trans Pattern Anal Mach Intel; 23(2):228-33, 2001.
[17] Martinez AM, Kak AC. PCA versus LDA. IEEE Trans Pattern Anal Mach Intel; 23(2):228-33, 2001.
[18] Cannady J. Artificial neural networks for misuse detection. National Information Systems Security Conference;1998. p. 368–81.
[19] J.R. Quinlan, C4.5 Programs for machine learning Morgan Kaufmann 1993.

AUTHORS PROFILE

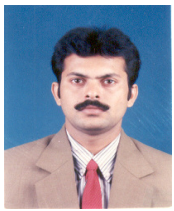 **Shailendra Singh** Lecturer in, Department of Information Technology at Rajiv Gandhi Technological University, Bhopal, India. He has publised Two research papers in International Journals and 8 papers in international and national conference proceedings His research interest include datamining and network security.He is a life member of ISTE, Associte member of Institution of Engineers (India) and member of International Association of Computer Science and Information Technology (IACSIT) Singapore.

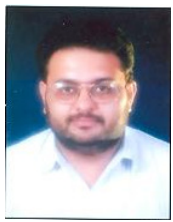 **Dr. Sanjay Silakari** Professor and Head, Department of Computer Science and Engineering at Rajiv Gandhi Technological University, Bhopal, India. He has awarded Ph.D. degree in Computer Science He posses more than 16 years of experience in teaching under-graduate and post-graduate classes. He has publised more than 55 papers in international and national journals and conference proceedings. He is member of International Association of Computer Science and Information Technology (IACSIT).